# Soft Computing in Product Recovery: A Survey Focusing on Remanufacturing System

Bo Xing[1], Wen-Jing Gao[1], Fulufhelo V. Nelwamondo[1], Kimberly Battle[1], and Tshilidzi Marwala[1]

[1]Faculty of Engineering and the Built Environment
University of Johannesburg
P. O. Box 524, Auckland Park 2006, Johannesburg, South Africa
{ bxing2009; wgao2011; vnelwamondo } @gmail.com, kbattle@uj.ac.za, tmarwala@uj.ac.za

**Abstract.** This paper focuses on the application of soft computing in remanufacturing system, in which end-of-life products are disassembled into basic components and then remanufactured for both economic and environmental reasons. The disassembly activities include disassembly sequencing and planning, while the remanufacturing process is composed of product design, production planning & scheduling, and inventory management. This paper presents a review of the related articles and suggests the corresponding further research directions.

## I. Introduction

Remanufacturing, or reman for short, in general, is a specific type of recycling in which end-of-life (EoL) products are reprocessed to a like-new condition. The goal of reman is to preserve most of the added-value by giving a second life to the EoL product and, meanwhile, to reduce the consumption of raw materials [1]. As we can see from the literatures that there is a fast growing body of papers. Among these papers, different techniques have been utilized. However there is a lack of survey available reviewing the various soft computing approaches applied to reman, exploring the current research trends and identifying opportunities for further research. This paper is trying to fill such gap by answering the following questions: what are the main problems within reman system have been investigated by using soft computing techniques? What kinds of techniques have been employed? The remainder of this paper is organized as follows: in Section 2, we will briefly overview the role of reman in the context of product recovery; following this section, the survey method employed in this research is introduced in Section 3; the results of this survey are detailed in Section 4 and the conclusions are drawn in the last section.

## II. The Role of Reman in Product Recovery

Briefly, product recovery aims to minimize the amount of waste sent to landfills by recovering materials and parts from old or outdated products by means of reprocess such as reman. Basically reman is an environmentally and economically sound way to achieve many of the goals of sustainable development: not only is it the case, as it is with other options e.g. recycling, that less waste must be landfilled and less virgin material consumed in manufacturing, but also the value added in the manufacturing of the components is also "recovered". It meanwhile saves the energy needed to transform and sort the material in recycling products. Consequently, reman can be seen as an advantageous product recovery option.

## III. Survey Methodology

The databases used in this study are provided by the library of University of Johannesburg, South Africa. During the study, the following online databases have been searched: ScienceDirect, Springer Link, Emerald, and Wiley Interscience. Currently this review covers only journal publications found from aforementioned databases. Two groups of keywords were used respectively to cross-search related journal papers in specific online databases. The key words used for soft computing are such as ant colony optimization (ACO), artificial immune system (AIS), artificial neural network (ANN), case based reasoning (CBR), Bayesian network, differential evolution (DE), evolution algorithm (EA), evolution programming (EP), fuzzy system (FS), genetic algorithm (GA), genetic programming (GP), greedy randomized adaptive search procedure (GRASP), memetic algorithm (MA), multi-agent system (MAS), neighborhood search (NS), particle swarm optimization (PSO), path relinking (PR), reinforcement learning (RL), scatter search (SS), simulated annealing (SA), and tabu search (TS). While the key words like disassembly, remanufacturing, end-of-life (EoL) product, product life cycle, and product recovery are used for reman system.

## IV. Results of Literature Survey on Reman System

The survey results are partitioned into the following two groups, namely disassembly and reman.

*A. Disassembly.* In the literature, disassembly is normally defined as a systematic method for separating a product into its constituent parts, components, subassemblies, or other groupings [2]. It is always considered to be one the most important steps in reman system. According to [3], sequencing and planning are two main activities involved in disassembly.

*1) Sequencing*: Disassembly sequencing deals with the problem of determining the best order of operations in the separation of a product into its constituent parts or other groupings [32]. The state-of-the-art approach for solving disassembly sequencing problem is TS. In [23], the authors employ TS to generate near optimal disassembly sequence.

However as shown in Table I, the most popular soft computing approach applied to this area is GA.

*2) Planning*: As pointed out in [3], disassembly sequencing addresses the question, "how to disassemble?" while disassembly planning delineates "how much to disassemble?" Generally there are two sub-problems involved in disassembly planning: disassembly-to-order (DTO) and line balancing. Not many soft computing methods have been used in this category and only ANN, FS, GA, and TS are found in the literature.

TABLE 1. Literature on disassembly

| Disassembly | | Soft Computing Approaches | | | | | | | | | | | | |
|---|---|---|---|---|---|---|---|---|---|---|---|---|---|---|
| | | *ACO* | *ANN* | *CBR* | *Bayesian Network* | *FS* | *GA* | *GP* | *GRASP* | *MAS* | *PR* | *RL* | *SS* | *TS* |
| Disassembly Sequencing | | [4] | [5] | [6] [7] | [8] | [4] [9] [10] | [11] [12] [13] [14] [15] [16] | [17] | [18] [19] [20] | [18] | [18] | [21] | [22] | [23] |
| Disassembly Planning | Line Balancing | [24] [25] [26] [27] | | | | | [27] [28] | | | | | | | |
| | DTO | | [29] | | | | [30] | [31] | | | | | | [23] |

**B. Reman.** According to the literature, there are various issues pertaining to reman. In this paper, we cover the following aspects: product design, production planning & scheduling, and inventory management.

*1) Product design*: Traditional product development aims at achieving improvements in design with respect to cost, functionality and manufacturability. However, increasing importance of the environmental issues forces product designers to consider certain environmental criteria in the design process.

*2) Production planning & scheduling* : Since the management of production planning and control activities can differ greatly from management activities in traditional manufacturing, the production planning is more complex for remanufacturing firms due to uncertainties from stochastic product returns, imbalances in return and demand rates, and the unknown condition of returned products. Accurate estimation of product returns is an important input for the analysis of remanufacturing systems.

*3) Inventory management*: Inventory management is a crucial area, especially when the decisions for the company in order to satisfy the following demands: (a) order of new items, (b) inspection of returned items which can be either as good as new or remanufacturable ones, and finally (c) remanufacturing of items that upon inspection are found to be remanufacturable. The goal is the minimization of the expected cost of the system in an infinite time horizon.

TABLE 2. Literature on reman

| Reman | Soft Computing Approaches | | |
|---|---|---|---|
| | ANN | FS | GA |
| Product Design | [33], [34], [35] | [36], [37], [38], [35] | [39], [34] |
| Production planning & scheduling | | [40], [41] | [42], [43] |
| Inventory management | | [44], [45] | [44] |

V. **Conclusions**

As show in Table 3, both GA and FS are the most popular approaches used in reman research. ACO has also been broadly employed in solving disassembly related problems. But there are some certain aspects in reman system have not yet been fully addressed by soft computing techniques. This is definitely a future direction for both research communities: soft computing and reman. Another observation is that most of the research papers employ more than one soft computing approach. From the soft computing point of view, although individual techniques have been applied successfully to solve many real-world problems, the current trend is to develop hybrids, since no one approach is superior to the others in all situations. Therefore, in terms of reman research, a future direction could be either improving the efficiency and effectiveness of existing methodologies or to create new hybrids that well suited for reman.

TABLE 3. Classification of publications based on soft computing approaches

| Soft Computing Approaches | References | No. of References |
|---|---|---|
| ACO | [4], [24], [25], [26], [27] | 6 |
| ANN | [5], [29], [33], [34] | 4 |
| Bayesian Network | [8] | 1 |
| CBR | [6], [7] | 2 |
| FS | [4], [9], [10], [30], [36], [37], [38], [40], [41], [44], [45], [35] | 12 |
| GA | [11], [12], [13], [14], [15], [16], [27], [28], [31], [39], [34], [42], [43], [44], [35] | 14 |
| GP | [17] | 1 |
| GRASP | [18], [19], [20] | 3 |
| MAS | [18] | 1 |
| PR | [18] | 1 |
| RL | [21] | 1 |
| SS | [22] | 1 |
| TS | [23] | 1 |

**Acknowledgment.** This work is partially supported by National Research Foundation (NRF), South Africa and Dean's Office, Faculty of Engineering and the Built Environment, University of Johannesburg, South Africa. The authors also gratefully acknowledge the helpful comments and suggestions of the reviewers, which have improved the presentation.